**Raman spectroscopic analysis of highly-concentrated antibodies under the acid-treated conditions**

Yusui Sato[1], Satoru Nagatoishi[2,3*], Shintaro Noguchi[4], Kouhei Tsumoto[2,3,5*]

[1] Analytical Instruments R&D Division, HORIBA, Ltd., Kanda Awaji-cho 2-6, Chiyoda-ku, Tokyo 101-0063, Japan

[2] The Institute of Medical Science, The University of Tokyo, 4-6-1, Shirokanedai, Minato-ku, Tokyo 108-8639, Japan

[3] Center for Drug Design Research, National Institutes of Biomedical Innovation, Health and Nutrition, 7-6-8 Saito-Asagi, Ibaraki City, Osaka 567-0085, Japan.

[4] Bio· Life Science Center, HORIBA, Ltd., 2 Miyanohigashi, Kisshoin, Minami-ku, Kyoto, 601-8510, Japan.

[5] Department of Bioengineering, School of Engineering, The University of Tokyo, 7-3-1, Hongo, Bunkyo-ku, Tokyo 113-8656, Japan.

* Correspondence and requests for materials should be addressed to N.S. (ngtoishi@ims.u-tokyo.ac.jp)

* Correspondence and requests for materials should be addressed to K.T. (tsumoto@bioeng.t.u-tokyo.ac.jp)

**Abstract**

Antibody drugs are usually formulated as highly-concentrated solutions, which would easily generate oligomer and aggregation, resulting in loss of efficacy. Although low pH increases the colloidal dispersion of antibodies, acid denaturation can be an issue. Therefore, knowing the physical properties of antibodies at low pH under high concentration conditions is an important insight into the quality evaluation of high concentration antibodies. Raman spectroscopy is a powerful tool to obtain conformational information derived from amino acid residues and secondary structures without dilution. In this study, Raman spectroscopy was used to investigate pH-induced conformational changes of antibodies at high concentrations. Raman experiments in pH 3 to 7 were performed for human serum IgG and recombinant rituximab. We detected the evident changes at pH 3 in Tyr and Trp Raman bands, which are the sensitive markers of intermolecular interactions. Thermal transition analysis over the pH range demonstrated that the transition temperature (aggregation temperature, $T_{agg}$) was highest at pH 3. Acid-treated and neutralized one showed higher $T_{agg}$ than that of pH 7, indicating that acid-induced conformational changes were not completely reversible. Colloidal analyses confirmed the findings of the Raman analysis, validating Raman spectroscopy as a method for evaluating antibody conformation associated with colloidal information.

**Introduction**

  Antibodies have become important biopharmaceuticals, especially in molecular target drugs, due to their great affinity with antigens and stability. Antibody drugs are usually formulated as highly-concentrated solutions. To establish the evaluation methods of antibody drugs during drug discovery, formulation, and production is a critical issue to improve process efficiency and manufacture safer products. A number of analytical techniques have been used to evaluate antibody function and stability during development [1–3]. Functional analyses are conducted based on antigen-antibody interactions using commercially available biosensors or antibody-dependent cellular cytotoxicity using cell-based assays. Stability assessment is necessary to ensure that the antibody does not form oligomers or aggregates and decreased drug efficacy under conditions used for formulation and storage. Various analytical methods are widely used to obtain colloidal and conformational information. However, many require that samples be diluted mainly due to the principal limitation prior to testing. Direct evaluation without dilution is essential for understanding the behavior of antibodies in highly-concentrated solution (non-ideal solution), because phenomena and theory in diluted solution (ideal solution) are quite different. Small-angle X-ray scattering (SAXS), small-angle neutron scattering, and static light scattering are available to evaluate highly-concentrated solutions of proteins [4–7]. These types of analyses do not provide information on individual functional groups and few conformational studies of highly-concentrated antibody solutions have been reported. Although high-concentration protein samples have been analyzed using nuclear magnetic resonance (NMR) [8,9], the method is still in its infancy for use as an antibody-drug development tool.

Raman spectroscopy is widely used to perform conformational analysis for proteins. Raman spectroscopy can provide conformational information derived from amino acid residues, especially aromatic ones and secondary structures as demonstrated by analyses of model proteins [10–17]. Spectral changes have been demonstrated to occur upon heating that result from alterations in protein conformation [12,13,15,16,18]. Our group carried out a study for high concentration solutions of an antibody [10]. Middle and short-range interactions are strengthened as protein concentration increases and distributions of these forces are shifted in each concentration. Tyr Raman bands are a sensitive indicator for hydrogen bonding and middle-range interactions [10]. Additionally, CH-π interactions representing short-range interactions occur in > 80 mg/ml solution as supported by changes in Raman bands of aromatic rings [10].

A low pH environment is chosen as one of the strategies to facilitate the stable dispersion of antibodies in aqueous solution. Under low pH conditions, the surface charge of the antibody is biased toward a positive charge, which increases colloidal stability due to electrostatic repulsion between antibodies. On the other hand, acid exposure is a severe issue in the production process of antibody pharmaceuticals. Although treatment with acidic solutions is one of the key steps used in the column purification of antibodies and inactivation of viruses, antibodies are always at risk of acid denaturation due to acid exposure [19,20]. Therefore, acid-induced effects on stability and conformation of antibodies are well studied [21–26]. Simple pH titration did not observe complete recovery to the native structure [21]. This result demonstrated that the acid stress, once given, was continuous. Acid exposure would generate aggregate form significantly, but the detailed investigation of the antibody conformation in a highly-concentrated solution is insufficient. Therefore, conformational analysis of acid-induced effects in a highly-concentrated antibody solution and the experimental proof associated with colloidal information, particularly about aggregate formation, is required.

In this study, we characterized acid-induced conformational changes in highly-concentrated antibody solutions. We performed a structural analysis using Raman spectroscopy under several pH conditions for two different antibodies. Commercially available human serum IgG (hIgG) and recombinant rituximab as a model of biopharmaceuticals were selected in this study. Raman spectral changes on heating were measured over a pH range from 3 to 7. We also monitored the effect of neutralization of the samples prepared at pH 3 and evaluated the acid-induced alterations. Dynamic light scattering (DLS) measurements for two antibodies were carried out to investigate the correlation between colloidal information and the conformational properties observed using Raman spectroscopy. Particle tracking analysis (PTA) of rituximab was conducted to quantify aggregation upon acid exposure and thermal stress. For rituximab, the effect of low pH on binding to its antigen was evaluated using a surface plasmon resonance imaging (SPRi) system.

**Materials and methods**

**Antibody preparation**

IgG from human serum (hIgG) was purchased from Sigma-Aldrich (I4506) and dissolved in 20 mM citrate-phosphate buffer at pH 3.0, 4.0, 5.0, 6.0, and 7.0. Recombinant rituximab was expressed and purified as described previously [27]. The expression vector was transiently transfected into ExpiCHO cells (Thermo Fisher Scientific) using ExpiFectamine CHO Transfection Kit (Thermo Fisher Scientific) following the manufacturer's recommended protocol. The cells were cultured for 8 days at 37ºC and 8% $CO_2$. The cultures were centrifuged at 400 *g* for 15 min, and the supernatant was collected. The supernatant was applied onto a rProtein A Sepharose Fast Flow column (GE Healthcare) equilibrated with PBS at pH 7.4. The fraction bound to the column was washed with the PBS and subsequently eluted with Pierce IgG Elution Buffer (Thermo Fisher Scientific). The eluted fraction was neutralized by the addition of 2 M Tris–HCl (pH 8.0) and further purified by size exclusion chromatography using a HiLoad 16/600 Superdex 200 pg column (GE Healthcare) equilibrated with PBS at pH 7.4. The solution of rituximab was enriched using an Amicon Ultra-4 50k centrifugal filter unit (Merck), and the buffer was exchanged for 20 mM citrate-phosphate at either pH 3.0 or pH 7.0. For acid-treated experiments, antibodies were incubated in pH 3.0 buffer for 1 h at room temperature. Then, the solutions were neutralized by addition of 1 M citrate-phosphate buffer at pH 8.0 and finally adjusted to pH 7.0 .

**Raman spectroscopic analysis**

Raman spectroscopy system was composed of a CCD camera (Synapse, HORIBA

Instruments Incorporated), a monochromator (MicroHR, HORIBA Instruments Incorporated) with a 1800 gr/mm grating (HORIBA France), and a 532 nm laser with a 100 mW power source (JUNO 532S, KYOCERA SOC Corporation). An optical microcell (M-30-G-5, GL Science) was used for all measurements. Solutions of antibodies at 50 mg/ml in 20 mM citrate-phosphate buffer at desired pH were heated from 25 to 90ºC at a temperature interval of 1ºC and the Raman spectra were collected with an acquisition time of 5 sec and 15 times accumulation at each temperature using Labspec 6 software (HORIBA France). The transition temperature (aggregation temperature, $T_{agg}$) was calculated by fitting with a sigmoidal function on OriginPro software (OriginLab Corporation).

**DLS experiments**

DLS measurements were performed using a nanoPartica SZ-100V2 (HORIBA, Ltd.). Solutions of 2 mg/ml antibody with or without heating at a given temperature for 30 min were placed in a quartz cell with an optical pass length of 10 mm. Light scattering was detected with a 532-nm laser with a power of 30 or 100 mW at a scattering angle of 90º. The data were recorded for 30 sec at 25ºC and the size distribution of the particles was calculated. The results were described as a relative scattering intensity graph.

**PTA experiments**

PTA measurements were carried out using ViewSizer 3000 (HORIBA Instruments Incorporated). A solution of 1 mg/ml rituximab with or without heating for 30 min was placed in a quartz cuvette with an optical pass length of 10 mm. Light scattering was detected with three lasers: a 445-nm laser with a power of 100 mW, a 520-nm laser with a power of 12 mW, and a 635-nm laser with a power of 8 mW. The scattering angle was 90º. For each solution, 25 videos were recorded at 25ºC with minimal stirring between each video to obtain fresh aliquot for respective data collection.

**SPRi experiments**

Antigen-antibody interactions were analyzed using an SPRi label-free interaction analysis platform, OpenPlex (HORIBA France). The antigen, CD20 (Sino Biological, 11007-H34E-B), was immobilized on the surface of a CS-HD biochip (HORIBA France) with the amine coupling method. Neutral and acid-treated rituximab solutions were diluted into 20 mM citrate-phosphate buffer at pH 7.0 (running buffer) to 5 nM. The association was recoded at a rate of 50 μl/min for 180 sec. The dissociation was measured by flowing running buffer alone for 540 sec, followed by injecting 100 mM Gly-HCl at pH 3.2 for 60 sec to regenerate the biochip. The experiments were performed at room temperature, and the reflectivity variation (%) in ligand spots was calculated by subtracting the signals of untreated areas on the chip surface. The collected data were analyzed using the ScrubberGen software (HORIBA France).

## Results

**Raman analysis of human serum IgG under acidic conditions**

We performed Raman spectral measurements of commercially available human polyclonal antibody IgG isolated from human serum (hIgG, the main component is IgG) in buffers of different pHs. The spectra from 550 to 1800 cm$^{-1}$ are shown in Fig. 1A. Several Raman bands typically observed in proteins were detected in all pH conditions. One of these is the band at 1670 cm$^{-1}$, which is assigned to the amide I band (C=O stretching of carboxyl groups) and is an indicator of the secondary structure of proteins. In our Raman measurements, no significant conformational changes were observed in the secondary structure of hIgG over the pH range evaluated. The bands of aromatic residues Phe at 1004 cm$^{-1}$ (ring breathing mode) and Trp at 1555 cm$^{-1}$ (C=C stretching vibration) are markers of tertiary structure [10]. Phe at 1004 cm$^{-1}$ was well discussed as a representative band of short-distance range interaction mainly due to the CH-π interaction. We noted the variation of the band width at 1555 cm$^{-1}$ to explain the molecular interactions and the excluded volume effect. Fermi doublets of Tyr at 830 and 850 cm$^{-1}$ are also referred to as a marker of tertiary structure. The intensity ratio of two peaks ($I_{850}/I_{830}$) is a marker of hydration and solvent exposure. This value is sensitive to Tyr side-chain interactions with neighboring molecules and is described as key bands in middle distance range interaction of highly-concentrated antibody solution [10]. In our experiments, only slight changes were observed at 1004 cm$^{-1}$ associated with pH variation. The values of the $I_{850}/I_{830}$ ratio and the band width at 1555 cm$^{-1}$ were compared in pH 3 to 7 and clearly decreased values in pH 3 were seen (Fig 1. B, C, D, E). These results suggest that the microenvironment of Tyr and Trp residues of hIgG under the acidic condition of pH 3 is different from the other pH conditions.

Thermal transition analysis of hIgG was carried out between 25 and 90°C in pH 3 to 7. The typical changes in Raman spectra associated with temperature increase were observed as shown in Fig. 2A, indicating that the signal intensity was decreased with heating, and the spectrum disappeared at the highest temperature. The transition curves of band intensity shifts of Phe at 1004 cm$^{-1}$, Trp at 1555 cm$^{-1}$, and amide I at 1670 cm$^{-1}$ were plotted (Fig. 2B). Each transition temperature (aggregation temperature, $T_{agg}$) was listed in Table 1. At each pH value, the $T_{agg}$ values of the three Raman bands were almost identical. The $T_{agg}$ value was highest at pH 3 of the conditions evaluated. To further assess the effect of pH on the stability of hIgG at pH 3, we incubated hIgG solution at pH 3.0 for 1 h and was neutralized prior to analysis. The thermal transition test was performed in the same manner as mentioned above (Fig. 2). The $T_{agg}$ values of the Raman bands at 1004 cm$^{-1}$, 1555 cm$^{-1}$, and 1670 cm$^{-1}$ were 66.3, 66.3, and 66.5 °C, respectively, resulting in similar to that of the pH 3 even though this measurement was conducted at neutral pH (Table 1). Thus, hIgG underwent a conformational change with increased $T_{agg}$ value at pH 3, and we found this conformational change was an irreversible transition.

**Raman analysis of rituximab under acidic conditions**

To further confirm the effect of pH 3 on highly-concentrated IgG antibody solutions, we prepared rituximab, a typical biologics of IgG, and evaluated its Raman spectra. Fig. 3A shows the Raman spectra of rituximab at pH 3 and 7. The $I_{850}/I_{830}$ ratio and the width of the band at 1555 cm$^{-1}$ at pH 3 were lower and narrower, respectively, than those at pH 7 (Fig.

3B, C, D, E), as was observed for hIgG. For rituximab at pH 7 the $T_{agg}$ values based on Raman bands at 1004 cm$^{-1}$, 1555 cm$^{-1}$, and 1670 cm$^{-1}$ were 69.3, 69.3, and 69.5 ºC, respectively (Fig. 4 and Table 2). These values are consistent with the melting temperature value measured with differential scanning calorimetry as reported in the previous research [28]. It is noteworthy that when the solution prepared at pH 3 was heated to 90ºC, only slight intensity shifts were observed over the entire spectrum. This result indicates that at pH 3, little or no aggregation occurred below 90°C. We also performed an acid-treated experiment using rituximab. The sample prepared at pH 3 was analyzed after neutralization. The thermal transition was clearly observed after neutralization, and each $T_{agg}$ value was slightly higher in the acid-treated than in the pH 7, indicating that the result of increased $T_{agg}$ in acid-treated one was consistent with hIgG (Fig. 4 and Table 2). Therefore, acid stimulation of human IgG with pH 3 induced an irreversible conformational change, which was clearly shown to improve the $T_{agg}$ value.

**Colloidal analyses of antibodies using DLS**

We performed a colloidal analysis of hIgG and rituximab using DLS to investigate aggregation tendencies associated with acid treatment and heating of antibody solutions (Fig. 5). The antibody solutions were heated at a temperature around $T_{agg}$ determined by Raman analysis (Table 1, 2). Monomeric state peaks of around 10 nm Stokes diameter were the main species observed in all pH conditions. In pH 7 and acid-treated samples, peak shifts and generation of aggregate form were observed upon heating, and large aggregated species of greater than 1000 nm Stokes diameter were detected at the boundary of $T_{agg}$. No oligomers or aggregates were detected in samples prepared at pH 3 with or without heating. Thus, DLS showed that the antibodies are colloidally stable in a pH 3 environment.

**Colloidal analysis of rituximab using PTA**

To evaluate the aggregation of antibody solutions quantitatively, PTA experiments were carried out using rituximab (Fig. 6). Heating treatment was given as the same manner as that of DLS measurements. The particle concentration was increased 61 fold upon the thermal stress at pH 7. In contrast, there was little change in particle number of antibodies at pH 3 when the sample was heated. The change in particle counts in these PTA analyses followed the same trend as the results of the DLS analyses. Interestingly, even in the acid-treated antibody solution, the difference in particle number with and without heating was also limited. These results suggested that acid-treated antibodies have altered physical properties and increased aggregation resistance.

**Binding activity of the acidic-treated antibody**

To examine whether acid treatment altered the function of rituximab, the binding of the antibody to antigen was evaluated using SPRi. A CD20 antigen was immobilized on a SPR sensor chip and rituximab was injected onto the sensor chip. The results are shown in Fig. 7. The shapes of binding curves of pH 7 and acid treated were similar with each other. However, the obtained signals of acid-treated were lower than that of pH7, suggesting that acid exposure caused irreversible conformational changes in rituximab that interfere with binding to antigen.

**Discussion**

We evaluated conformational variation of highly-concentrated antibodies, associated with pH change and heating using Raman spectroscopy. The polyclonal hIgG has been used in previous research, and its protein-protein interactions have been characterized in detail [10]. Rituximab is a representative monoclonal antibody of biopharmaceuticals with well-studied physicochemical properties [27–31]. We expected that the analysis of these two antibodies by using Raman spectroscopy would provide information relevant to other antibodies.

A detailed observation of the Raman spectra associated with the pH change revealed significant differences in Raman bands derived from aromatic amino acids. The band intensity ratio of Tyr ($I_{850}/I_{830}$) of hIgG in pH 3 was clearly lower than those in pH 4 to 7 (Fig. 1B, D). This indicates that the local environment of Tyr side chains, many of which are located in the Fab region, are different, and intermolecular interactions are weakened at acidic pH. This trend was consistent with the result of rituximab (Fig. 3B, D). A similar propensity was reported previously, which might be indicative of a general response under the acidic conditions [14]. Further, the band width of Trp at 1555 cm$^{-1}$ of hIgG in pH 3 was narrower than that of pH 4 to 7 (Fig. 1C, E), indicating that intermolecular interactions associated with Trp were suppressed in the low pH. This result was also seen in rituximab (Fig. 3C, E).

In acid-treated hIgG, the $I_{850}/I_{830}$ ratio was lower than that of pH 7 even though two measurements were carried out in the same neutral pH (Fig. 1B, D). This suggested that the effect of acid on the conformation of hIgG was irreversible. In contrast, almost similar values of the ratios were shown in acid-treated and pH 7 of rituximab (Fig. 3B, D). The band width at 1555 cm$^{-1}$ of acid-treated was close to that of pH 7 in both hIgG and rituximab (Fig. 1B, D, 3B, D). The different impacts of acid-stress on the two antibodies must arise from differences in stabilities of the Fab regions, and not the Fc region, which is considered to be a nearly identical amino acid sequence. Because rituximab has a fine-tuned structure as a biopharmaceutical, its acid resistance must be superior to hIgG. These findings were supported by thermal transition analysis and DLS measurements (Fig. 2, 4, 5, Table 1, 2). Our experiments demonstrated that the conformational changes in the acid stress were clearly detected in highly-concentrated antibody solutions and the initial structure was not fully recovered even after neutralization. The partially irreversible impacts on antibody structure was previously reported in analysis of dilute solutions [21]. In this paper, the condition at pH 2.7 showed the conformational change, but the effect of pH 3.5 was limited. This borderline would be consistent with results that the Raman spectra in pH 4 did not show the certain difference with pH 5 to 7, but pH 3 did especially in the bands of Tyr and Trp (Fig. 1, 3). Therefore, it is suggested that our observation with Raman spectroscopy would be well supported by the existing knowledge and remarkable from the viewpoint of conformational insights into high concentration antibodies. For rituximab, the effect of acid treatment on its ability of antigen binding was evaluated by SPRi measurements (Fig. 7). The decreased signals were detected in the acid-treated sample, suggesting that the exposure to low pH induced the loss of the antigen-binding activity due to the irreversible conformational changes in the Fab region.

For thermal transition analysis in Raman spectroscopy using hIgG and rituximab associated with pH variation (Fig. 2. 4), the decreased signals were detected upon the temperature increasing and transition temperatures ($T_{agg}$) of three Raman bands of Phe, Trp, and amide I related to the secondary and tertiary structures were calculated (Table 1, 2). The heating tests were also conducted in several studies using Raman spectroscopy and the decreased signals were observed with thermal stress and conformational changes of model proteins were discussed [12,13,15,16]. The $T_{agg}$ derived from the respective bands showed almost similar values, indicating that the significant differences were not observed in these experiments. In comparison of pH conditions, the highest $T_{agg}$ value was given in pH 3 and the comparable transition temperature was obtained in acid-treated in hIgG (Fig 2, Table 1). For rituximab, the only slight transition was detected in pH 3 and $T_{agg}$ in pH 7 and acid-treated were higher than those of hIgG (Fig. 2, 4, Table 1,2). This result would represent the clear effect of pH and intrinsic stabilities of hIgG and rituximab mainly due to Fab regions, as mentioned above.

From the obtained transition temperature, the thermal stress was given at around $T_{agg}$ values and then a particle size analysis using DLS was performed (Fig. 5). The oligomer or aggregate form was mainly detected at around $T_{agg}$, but not in pH 3. This result suggested that the $T_{agg}$ from Raman is well correlated with protein aggregation in a highly-concentrated solution. Actually, the transparency of the sample solution was completely lost after heating tests except for rituximab in pH 3. PTA experiment was also supportive of the pH effect on aggregate formation for rituximab (Fig. 6). PTA tracks individual particles directly unlike DLS, which enables to detect aggregates sensitively and characterize quantitatively especially in biologics with improved stability. PTA showed a quantitative difference in the counted particle number in pH 7 after heating and only slight changes in pH 3 and acid-treated. Aggregation propensity in acidic pH has been well studied for a diluted solution and low pH would easily induce aggregate form. A previous study has reported the aggregation and denaturation properties of individual domains of antibodies [25]. The acidic condition, pH 2-3 without NaCl, does not necessarily lead to the formation of aggregates. For CH2, CH3 domains and CH1-CL dimer, each fragment shows a monomeric monodispersed non-native state under the condition of pH 2-3 without NaCl. This phenomenon is consistent with our results of DLS and Raman experiments especially in the increased $T_{agg}$ in pH 3 (Fig. 2, 4, 5, Table 1, 2) and also supports the description that aggregation propensity predominantly depends on individual Fab regions. The colloidal analyses associated with pH variation in this study showed a positive correlation with the results of Raman experiments, suggesting that Raman spectroscopy could depict the aggregation properties of antibodies by observing the Raman bands such as aromatic residues.

In the studies of the highly-concentrated solutions, several experiments have been carried out using various analytical techniques. The excluded volume effect was discussed in Raman spectroscopy, and 50 mg/ml targeted in this study applies to this effect with the marker of Trp at 1555 cm$^{-1}$ [10]. Decreased the band widths of pH 3 in both hIgG and rituximab were shown (Fig. 1, 3), suggesting that low pH might weaken the intermolecular interactions and the excluded-volume effect. This assumption is in good agreement with the increase of $T_{agg}$ and the propensity of less aggregate formation in pH 3 (Table 1, 2). SAXS analysis is also a direct method to examine concentrated solutions. The increase of repulsive interactions with increasing protein concentration is shown using model antibodies with or without additives [32,33]. Rheological analysis and theoretical calculation also demonstrate the increase of the charge-charge repulsion effect due to the high concentration and low pH [34,35]. Viscoelastic properties show a good correlation with the protein-protein interactions and the modeling indicates that high concentration contributed to not only dipole-dipole attractions but also increase in net charge, resulting in charge-charge repulsions in pH 3-4. Thus, it is suggested that the suppressed intermolecular interactions should be attributed to electrostatic repulsions and supports our results of Raman experiments, which confirmed notable changes in bands of aromatic residues and the increased $T_{agg}$ in pH 3.

**Conclusions**

In this study, Raman spectroscopy was used to reveal information about conformational changes and intermolecular interactions in highly-concentrated solutions of antibodies as a function of pH. The ratio of intensities of band due to Tyr ($I_{850}/I_{830}$) and the width of the Trp band at 1555 cm$^{-1}$ were shown to be markers of protein-protein interactions in both a commercially available antibody and a model biopharmaceutics. Acid-induced irreversible effects were clearly shown by thermal transition tests. The results showed a good agreement with the previous experiments with dilute samples and presented the novel observation using highly-concentrated solutions associated with acid-treatment. Some attractive forces remain in concentrated solutions; charge-charge repulsions would be dominant and decrease the intermolecular interactions in acidic pH. The colloidal information obtained from DLS and PTA analyses supported the conclusions drawn from analyses of Raman spectra. Our work suggests that Raman spectroscopy provides unique information that cannot be obtained using other biophysical techniques regarding intermolecular interactions in highly-concentrated solution significantly linked to the colloidal properties and that this analytical method could be a useful technique for prediction of aggregation resistance and optimization of formulations. Comprehensive evaluation using biopharmaceuticals would improve the reliability of Raman spectroscopy as a tool for drug selection or quality control.

**Acknowledgements**

The authors thank Ms. Chizuru Saruta and Yumiko Yamagami for preparing recombinant rituximab.

**Authors' contributions**

Y.S., S. Nagatoishi, and K.T. conceived the study and designed experiments. Y.S. performed the Raman, DLS, PTA, and SPR analyses. Y.S., S. Nagatoishi, and K.T. wrote, and S. Noguchi assisted the manuscript. All authors approved the manuscript.

**Table 1.** Thermal transition temperatures of hIgG as a function of pH.[a]

|  | $T_{agg}$ at 1004 cm$^{-1}$ (°C) | $T_{agg}$ at 1555 cm$^{-1}$ (°C) | $T_{agg}$ at 1670 cm$^{-1}$ (°C) |
|---|---|---|---|
| pH 3.0 | 67.0 ± 1.9 | 66.9 ± 1.8 | 67.2 ± 1.9 |
| pH 4.0 | 61.5 ± 0.7 | 61.5 ± 0.7 | 61.5 ± 0.7 |
| pH 5.0 | 61.1 ± 0.1 | 61.1 ± 0.1 | 61.2 ± 0.1 |
| pH 6.0 | 61.4 ± 0.7 | 61.4 ± 0.7 | 61.4 ± 0.8 |
| pH 7.0 | 61.5 ± 1.2 | 61.5 ± 1.3 | 61.5 ± 1.2 |
| Acid-treated[b] | 66.3 ± 0.1 | 66.3 ± 0.2 | 66.5 ± 0.1 |

[a] Values are means ± standard deviations derived from three independent measurements. [b] The sample was incubated at pH 3 for 1 h and neutralized prior to measurement.

**Table 2.** Thermal transition temperatures of rituximab as a function of pH.[a]

|  | $T_{agg}$ at 1004 cm$^{-1}$ (°C) | $T_{agg}$ at 1555 cm$^{-1}$ (°C) | $T_{agg}$ at 1670 cm$^{-1}$ (°C) |
|---|---|---|---|
| pH 3.0 | - | - | - |
| pH 7.0 | 69.3 ± 0.3 | 69.3 ± 0.3 | 69.5 ± 0.2 |
| Acid-treated[b] | 71.1 ± 0.1 | 71.1 ± 0.1 | 71.2 ± 0.1 |

[a] Values are means ± standard deviations derived from three independent measurements. [b] The sample was incubated at pH 3 for 1 h and neutralized prior to measurement.

**Figure legend**

**Fig. 1.** Raman spectra of 50 mg/ml hIgG in buffer of pH 3.0 (blue), 4.0 (orange), 5.0 (green), 6.0 (pink), 7.0 (black) or acid-treated (gray). Signals from appropriate buffer were subtracted from intensities of samples. (A) Raman spectra in the range from 550 to 1800 cm$^{-1}$. (B) Tyr Raman bands at 830 and 850 cm$^{-1}$. (C) Trp Raman bands at 1555 cm$^{-1}$. (D) Plots of the band intensity ratio of $I_{850}/I_{830}$ and (E) band width of Trp at 1555 cm$^{-1}$ as a function of pH. Open circle represents the acid-treated condition. Values are means ± standard deviations derived from three independent measurements.

**Fig. 2.** Thermal transition experiments of hIgG using Raman. (A) Typical Raman spectra changes by heating. Intensities without blank subtractions were decreased associated with temperature increase in the range from 25 to 90 ºC. (B-D) Plots of the normalized band intensity of Phe at 1004 cm$^{-1}$, (C) Trp at 1555 cm$^{-1}$, and (D) amide I at 1670 cm$^{-1}$ bands in pH 3.0 (white diamond), pH 4.0 (gray diamond), pH 5.0 (white square), pH 6.0 (gray square), pH 7.0 (white circle), and under acid-treated conditions (gray circle) as a function of temperature.

**Fig. 3.** Raman spectra of 50 mg/ml rituximab in solution of pH 3.0 (blue), 7.0 (black), and acid-treated (gray). Signals from appropriate buffer were subtracted from intensities of samples. (A) Raman spectra in the range from 550 to 1800 cm$^{-1}$. (B) Tyr Raman bands at 830 and 850 cm$^{-1}$. (C) Trp Raman bands at 1555 cm$^{-1}$. (D) Plots of the band intensity ratio of $I_{850}/I_{830}$ and (E) band width of Trp at 1555 cm$^{-1}$ as a function of pH. Open circle represents the acid-treated condition. Values are means ± standard deviations derived from three independent measurements.

**Fig. 4.** Thermal transition experiments of rituximab using Raman in the range from 25 to 90°C. (A) Plots of the normalized band intensity of Phe at 1004 cm$^{-1}$, (B) Trp at 1555 cm$^{-1}$, and (C) amide I at 1670 cm$^{-1}$ bands in pH 3.0 (white diamond), pH 7.0 (white circle), and acid-treated (gray circle) as a function of temperature.

**Fig. 5.** DLS measurements of hIgG and rituximab. (A) hIgG in pH 3.0, (B) pH 7.0, (C) acid-treated, (D) rituximab in pH 3.0, (E) pH 7.0, and (F) acid-treated with heating incubation below the $T_{agg}$ (black broken line), above the $T_{agg}$ (gray line), or without heating (black line).

**Fig. 6.** PTA experiments of rituximab. (A) pH 3.0, (B) pH 7.0, and (C) acid-treated with heating treatment at 67°C (black bar) or without heating (gray bar).

**Fig. 7.** Antigen binding analysis. CD-20 was immobilized on a biochip and 5 nM rituximab prepared at pH 7.0 (black line) or acid-treated (broken line) was flowed over the chip surface at rate of 50 μl/ml for 180 sec, followed by 540 sec dissociation.

**Fig. 1**

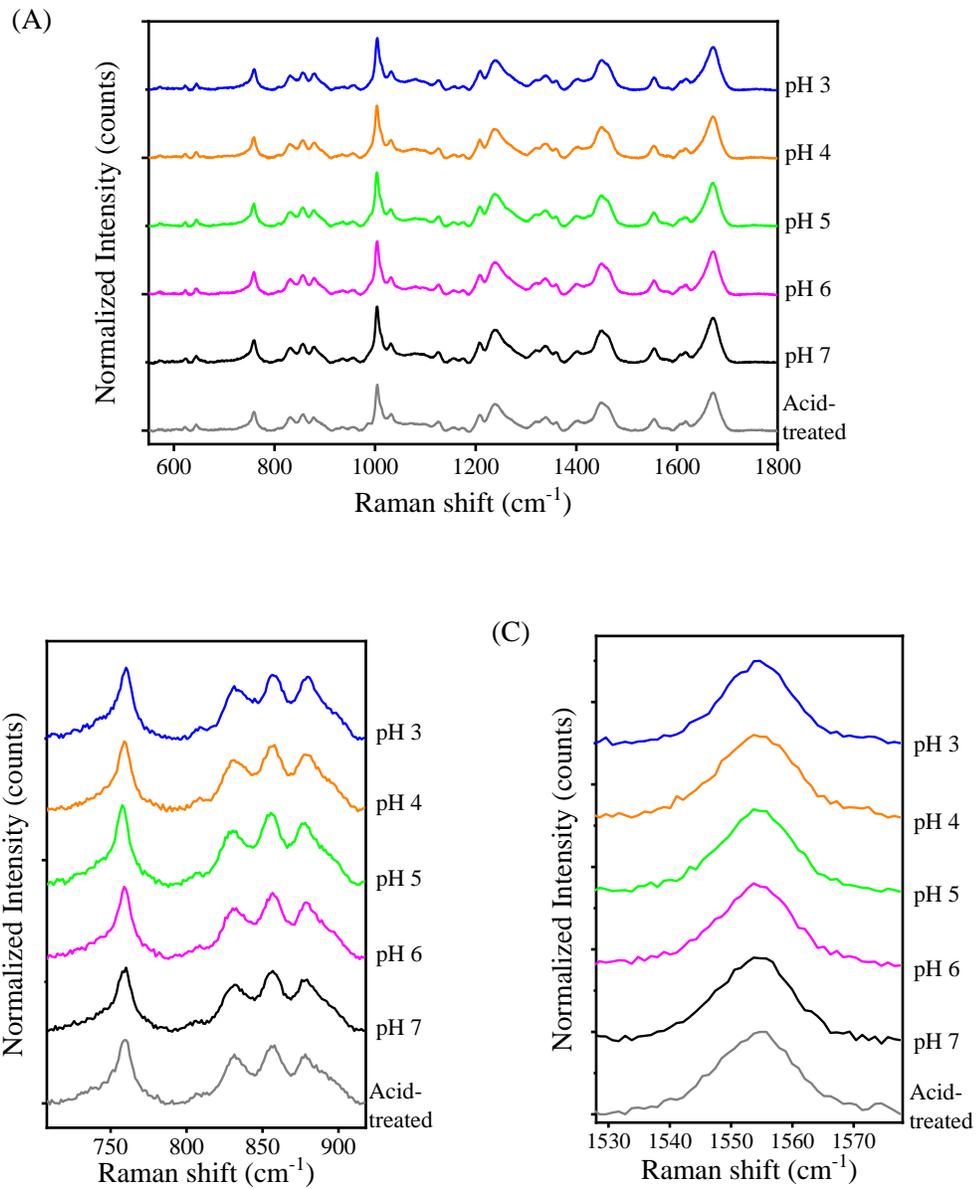
29

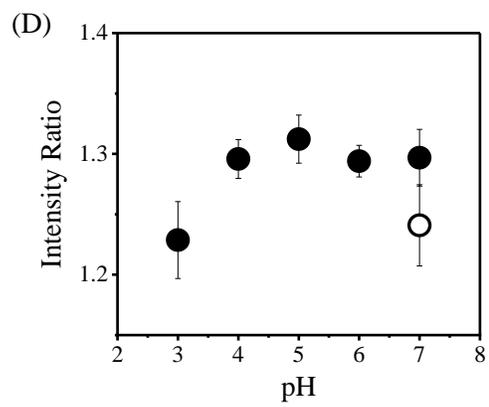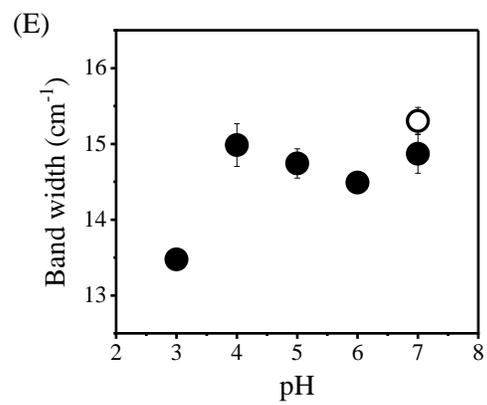

**Fig. 2**

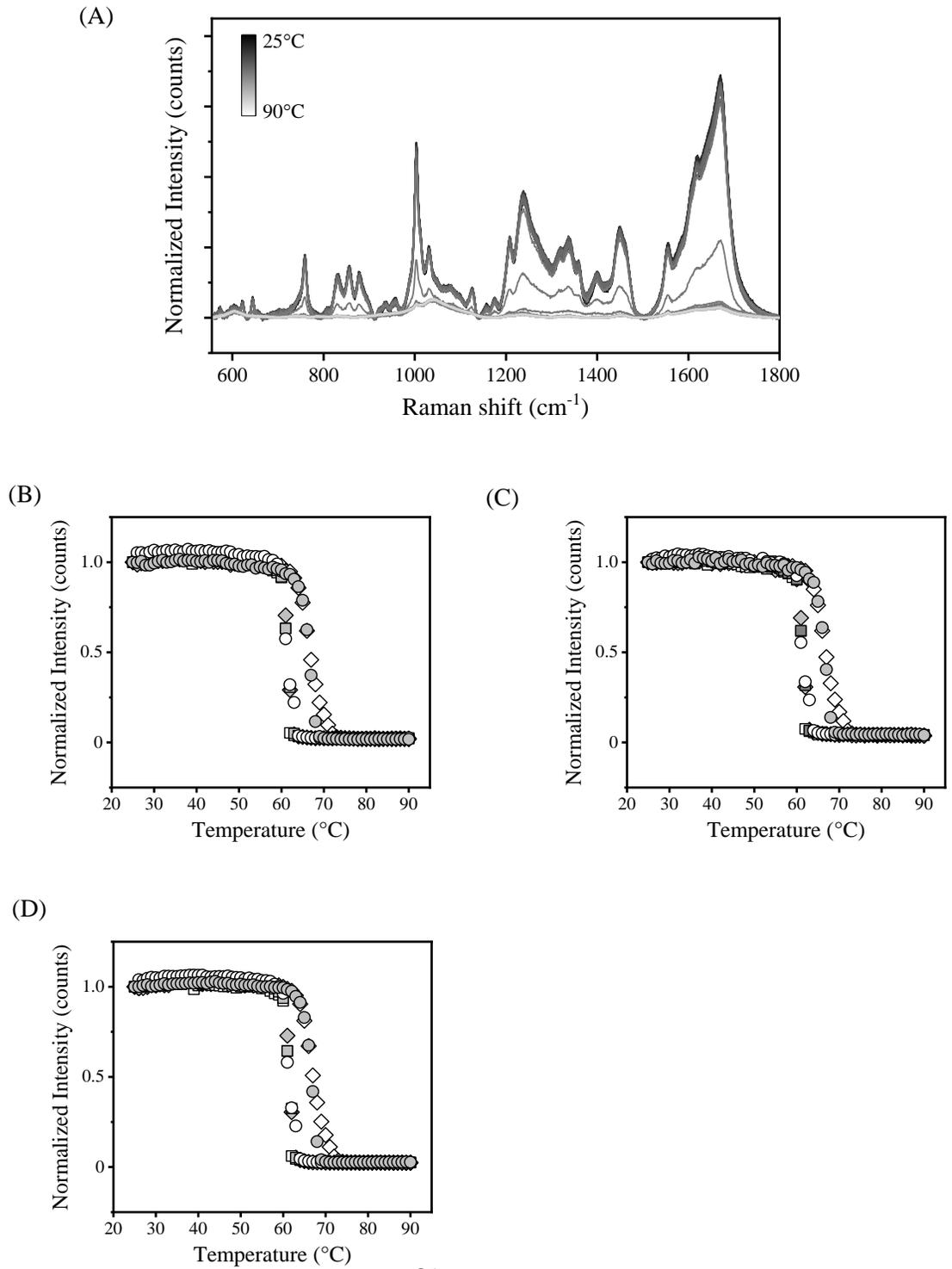

**Fig. 3**

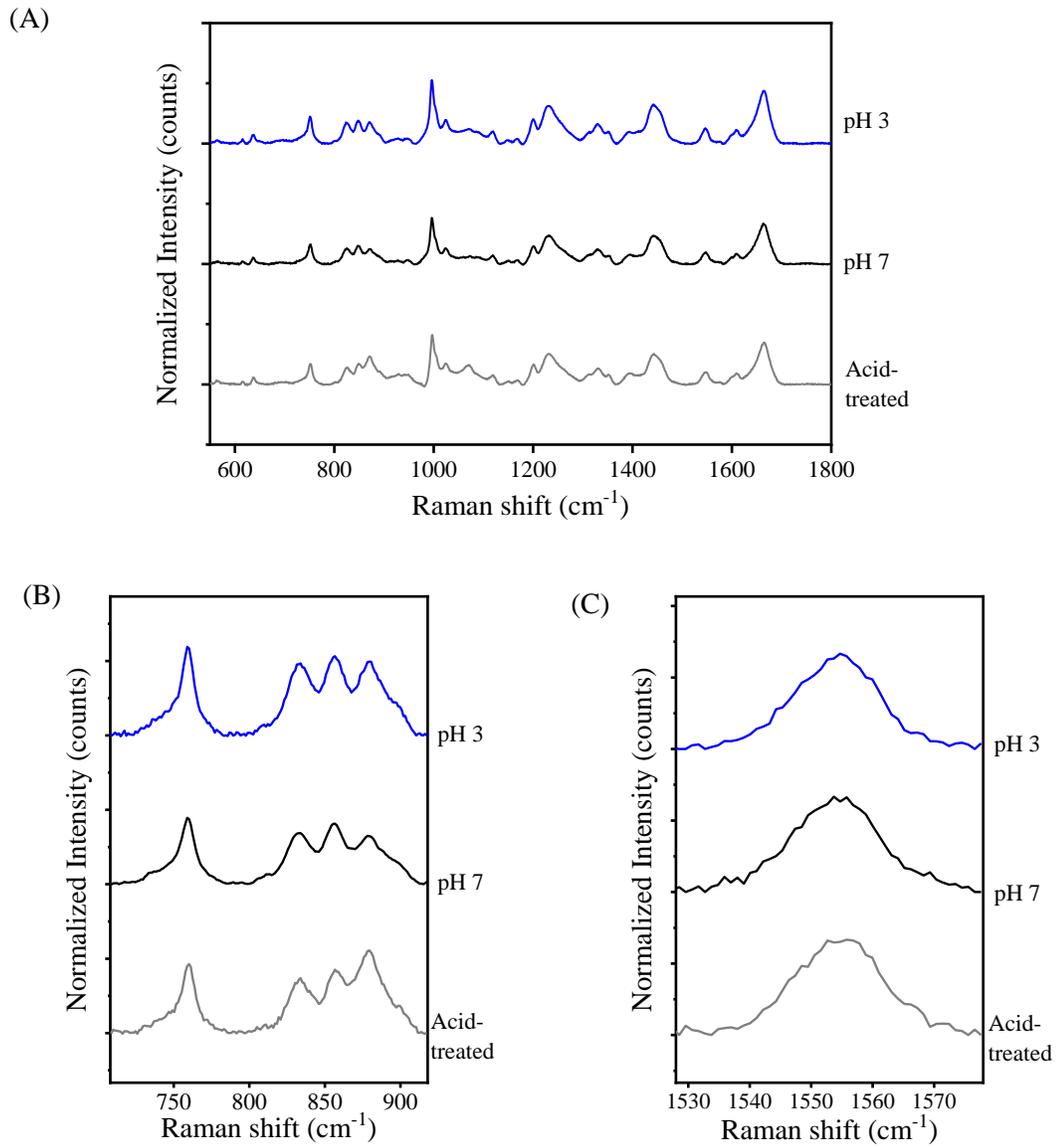

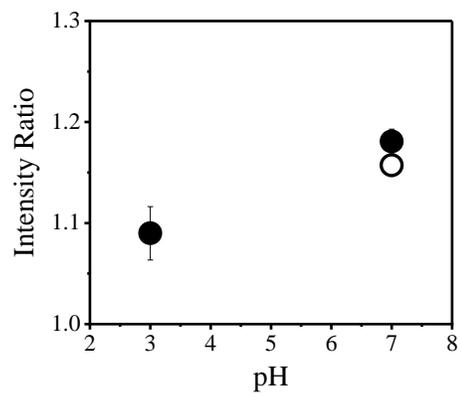 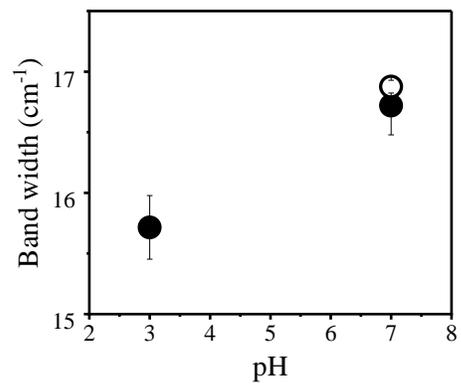

**Fig. 4**

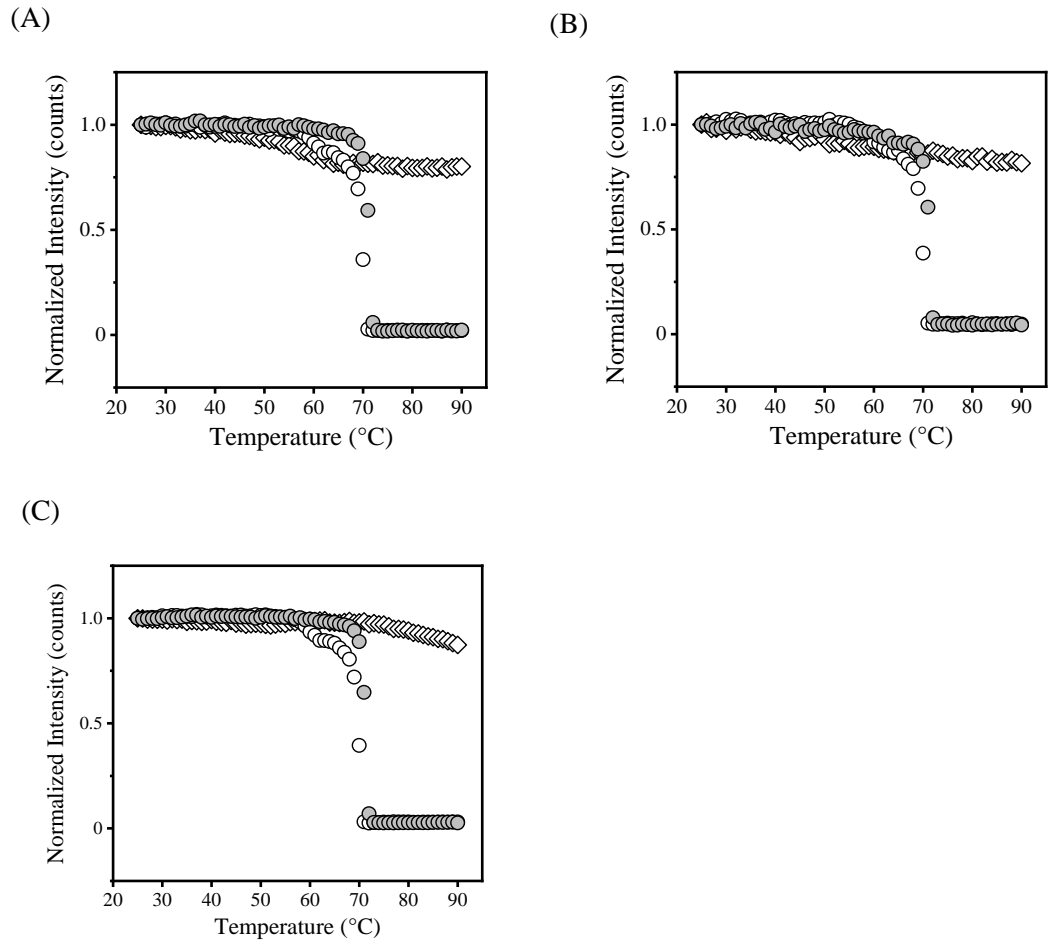

**Fig. 5**

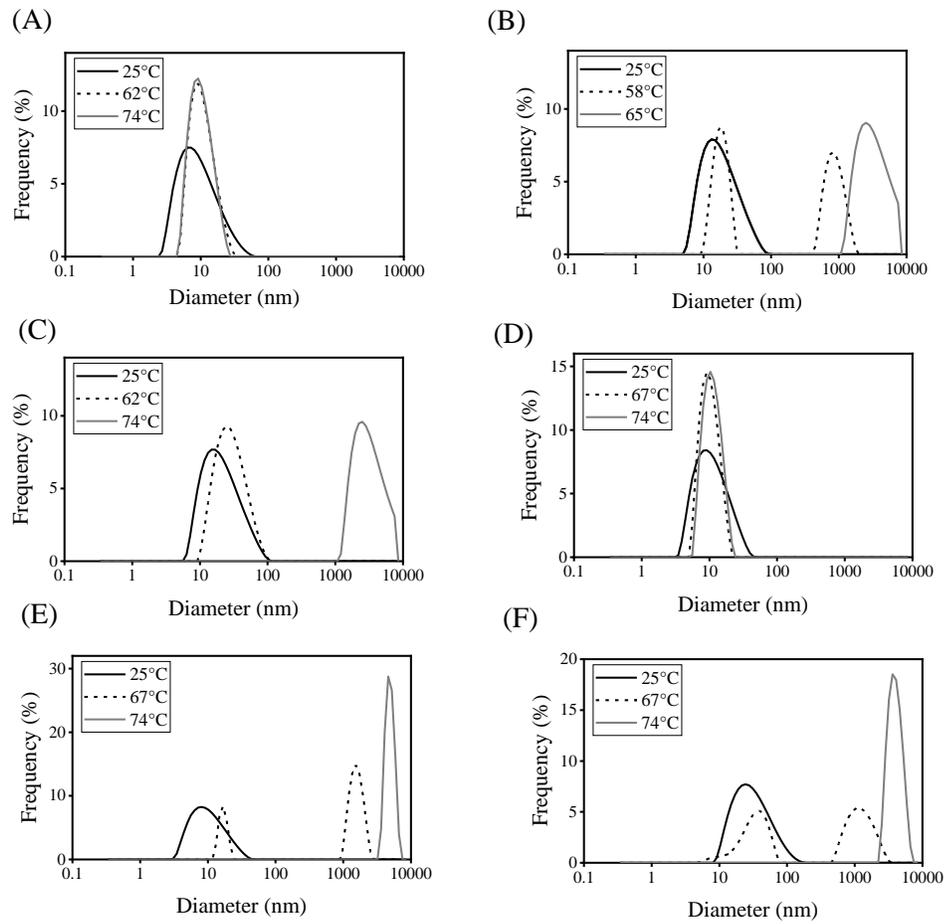

**Fig. 6**

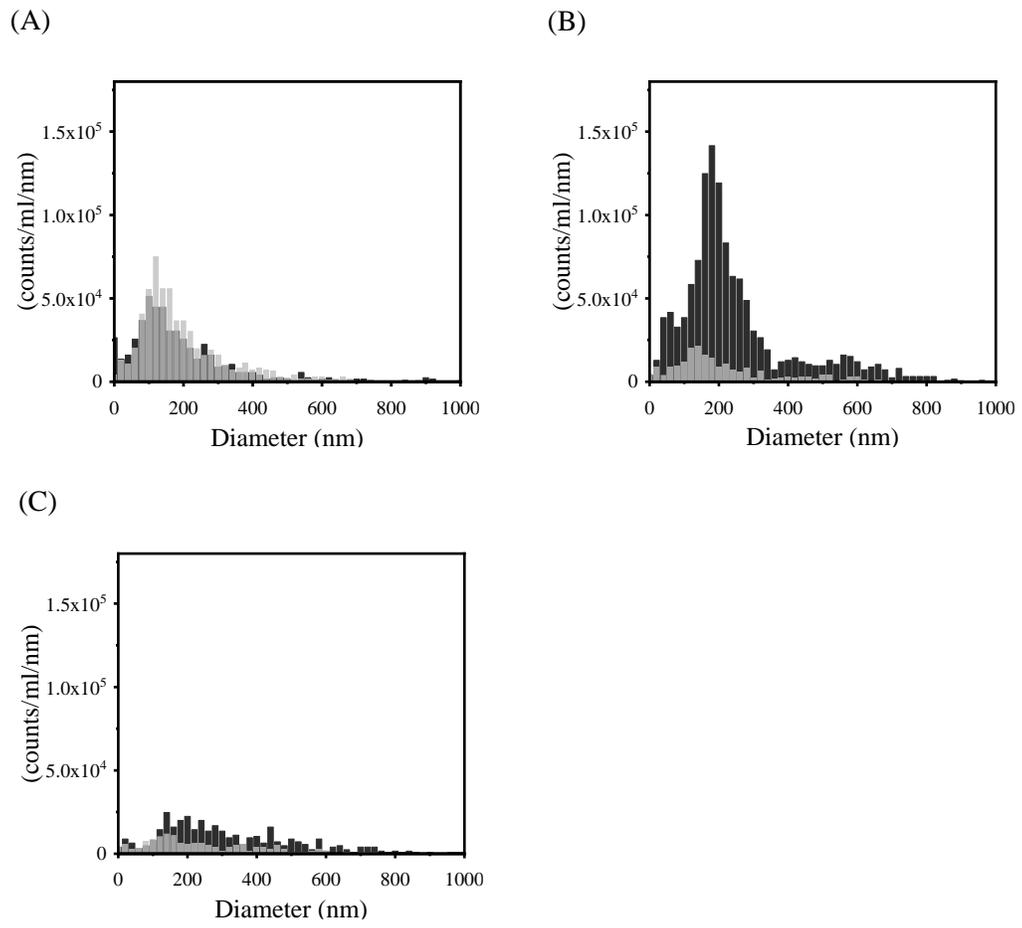

**Fig. 7**

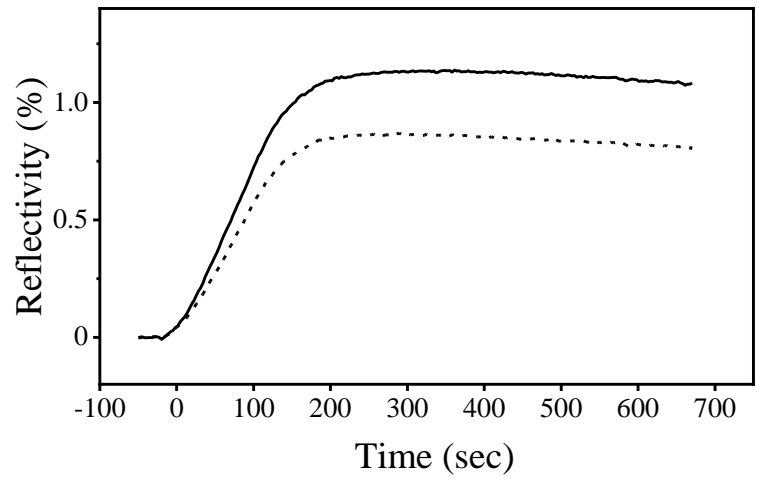